# Local solid-state modification of nanopore surface charges

**Ronald Kox**[1,2,5], **Stella Deheryan**[1], **Chang Chen**[1,3], **Nima Arjmandi**[1], **Liesbet Lagae**[1,4], and **Gustaaf Borghs**[1,4]

Imec, Kapeldreef 75, 3001, Leuven, Belgium

Department of Electrical Engineering, Katholieke Universiteit Leuven, Kasteelpark Arenberg 10, 3001, Leuven, Belgium

Department of Chemistry, Katholieke Universiteit Leuven, Celestijnenlaan 200 F, Leuven, 3001, Leuven, Belgium

Department of Physics, Katholieke Universiteit Leuven, Celestijnenlaan 200 D, Leuven, 3001, Leuven, Belgium

**Abstract**: The last decade, nanopores have emerged as a new and interesting tool for the study of biological macromolecules like proteins and DNA. While biological pores, especially alpha-hemolysin, have been promising for the detection of DNA, their poor chemical stability limits their use. For this reason, researchers are trying to mimic their behaviour using more stable, solid-state nanopores. The most successful tools to fabricate such nanopores use high energy electron or ions beams to drill or reshape holes in very thin membranes. While the resolution of these methods can be very good, they require tools that are not commonly available and tend to damage and charge the nanopore surface. In this work, we show nanopores that have been fabricated using standard micromachning techniques together with EBID, and present a simple model that is used to estimate the surface charge. The results show that EBID with a silicon oxide precursor can be used to tune the nanopore surface and that the surface charge is stable over a wide range of concentrations.

## 1 Introduction

Examples of nanopores are abundant in biological systems, usually in the form of transmembrane protein channels in lipid bilayer membranes. While is it possible to use these biological nanopores for the characterization of biomolecules like DNA, their poor stability limits their use outside of the lab [12, 5]. Fortunately, the ever continuing trend of downscaling has resulted in functional components with sizes comparable to individual biomolecules, making it possible to fabricate more stable, artificial nanopores. The biosensing principle of nanopores typically depends on a change of a property like ionic conductance or induced charges during the translocation of a biomolecule [27, 10, 23].

Over the years, several approaches have been used to fabricate nanopores, usually employing some sort of feedback system to control the size [17, 26, 18, 2, 11, 16, 13, 19]. One of the most successful methods to date uses the high energy electron beam of a transmission electron microscope (TEM) to fine-tune the size of an existing nanopore with nanometre precision and visual feedback, in both silicon nitride and silicon oxide membranes [26, 16]. It is also possible to directly drill nanopores using a TEM or a focused ion beam (FIB) [17, 13]. While these methods can accurately control the size, they offer little control over the resulting surface properties of the nanopore because of damage and charge implantation resulting from the ion or electron beam.[14] Some attempts have been made to solve these problems by changing the surface using atomic layer deposition or chemical modifications [4, 29].

---

[5] Corresponding author: ronald.kox@imec.be

In this work, the nanopores were fabricated using a combination of microfabrication and electron-beam induced deposition (EBID). EBID is a direct-write nanofabrication process that involves the local deposition of a solid material onto a substrate by means of an electron-mediated decomposition of a precursor molecule, typically a vapour [22]. It has already been shown that it can be used for the fabrication of nanostructures, the precise deposition of small quantum dots and the extremely localized deposition of Ramann sensitive materials [28, 7, 3].

This deposition technique has also been successfully employed for the fine-tuning of existing nanopores, both for FIB drilled silicon nitride and micromachined silicon nanopores [6, 15]. While in this case the resolution of the resulting nanopores is limited by the resolution of the used SEM tool, the use of different precursors can give a much better control over the resulting surface properties. Moreover, since the acceleration voltage of a SEM is much lower than that of a TEM, charge implantation is much lower. The introduction of a precursor typically requires a gas injection system, but in this work we have used a special sealing device to introduce a silicon oxide precursor in an unmodified SEM. The resulting nanopores are chemically stable, and ionic measurement results show a constant surface charge over a wide range of salt concentrations. A simple analytical model can then be used to estimate the nanopore surface charge, and evaluate the influence of the EBID process.

## 2   Experimental Section

### 2.1   Nanopore fabrication

The nanopores described in this work are created using standard micro-machining techniques, similar to the method described in our previous work [15]. The wafers used for the fabrication were silicon-on-insulator (SOI) wafers from Soitec (France), with a 705 nm top silicon layer, a 1 µm buried oxide (BOX) and a 725 µm silicon substrate. The wafer was coated with 150 nm low pressure chemical vapor deposition (LPCVD) silicon nitride on the backside, and a 30 nm layer of high temperature plasma-enhanced chemical vapour deposition (PECVD) silicon oxide on the front side. Freestanding membranes could be formed by patterning square windows of about 1 mm x 1 mm in the silicon nitride on the backside using reactive ion etching (RIE) and a long, self-limiting anisotropic wet etch (12 hours, using 10% KOH at 80 °C), where the nitride served as a hard mask and the BOX as a stopping layer. Next, e-beam lithography (EBL) and buffered hydrofluoric acid (BHF) were used to form square windows of about 1 µm x 1 µm in the top silicon oxide layer. This layer, together with the BOX, is used for another, much shorter self-limiting anisotropic wet etch (8 minutes using 33% KOH at 40 °C). To open the resulting nanopores, the samples were submerged in BHF for a few minutes to remove the BOX. Finally, wet thermal oxidation was performed to grow at thin silicon oxide layer using a rapid thermal annealing oven (Heatpulse 310, AG-RTP, USA) with an $H_2O$ saturated nitrogen environment. The resulting nanopores have a pyramid shape, and the size of the opening is typically around 100 nm.

## 2.2 Electron-beam induced shrinking

A special sealing device for the introduction of a precursor into the SEM was developed, eliminating the need to install an expensive gas injection system. The sealing device is shown in figure 1, and it consists of three important parts: an aluminium casing with a reservoir, a chip with one or several nanopores, and PDMS seals. A liquid precursor can be introduced into the reservoir, which is then sealed off from the environment using the PDMS seals. When this device is introduced into the SEM, the nanopores in the chip provide a way for the precursor to slowly leak out into the chamber, providing a steady flow of precursor, which is slow enough for the SEM to reach high vacuum ($10^{-5}$ mbar or better) [9]. The SEM used in the experiments was a XL30 FESEM (Philips, The Netherlands), with an acceleration voltage of 5 kV and an emission current of 190 µA. The precursor that was introduced was tetramethoxysilane (TMOS), which is a precursor for the deposition of silicon oxide [20].

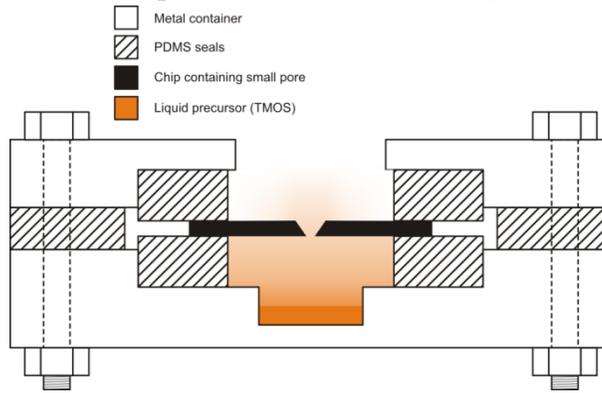

**Figure 1** Metal sealing device used for releasing precursor molecules into the chamber of a SEM. A small metal container, which can contain a liquid precursor like TMOS, is sealed from the environment by a chip containing a small hole, and several PDMS seals. When introduced into a SEM, the precursor will slowly evaporate through the nanopore.

## 2.3 Ionic measurements:

The measurement setup consists of two Plexiglas reservoirs, two PDMS seals, and a chip containing a nanopore mounted in between. Before mounting a chip in the measurement setup, it was first treated with oxygen plasma or UV ozone for several minutes, to increase hydrophilicity and improve wettability. Both reservoirs were then filled with a saline KCl solution with different concentrations (from 0.1 mM to 1 M). Each reservoir contained a Ag/AgCl electrode, which was connected to a current amplifier (Keithley 428-PROG, Keithley Instruments Inc, USA) to apply a potential bias and measure the ionic current through the nanopore. The ionic signal was recorded using a data acquisition card (NI PCIe-6363, National Instruments, USA) and analyzed using Matlab (R2009a, Mathworks, USA).

## 2.4 Analytical model:

To get an estimate of the surface charge present in the nanopore, we fitted our data to an analytical model, which is based on the combination of the Poisson Nernst-Planck equations, and the Donnan equilibrium as boundary conditions. The model is similar to the one used by Cervera et al., but it assumes a cylindrical nanopore instead of a conical one [1]. Assuming a fixed surface charge, a cylindrical nanopore, and Donnan equilibrium boundary conditions, the following expression was derived for the nanopore conductance.

$$G = -2\pi r_p^2 \left( \sum_i z_i c_i D_i \right) \frac{1}{L} \frac{F^2}{RT} \quad (i=\text{K, Cl}) \tag{1}$$

In this equation, $G$ is the nanopore conductance, $r_p$ is the nanopore radius, $L$ is the length of the pore (or membrane thickness), $z_i$ is the ion valence, $D_i$ is the diffusion constant for each ion, $T$ is the temperature, $F$ is the Faraday constant and $R$ is the gas constant.

The concentration $c_i$ for each ion is given by the following equation:

$$c_i = 1/2\left(-z_i X + \sqrt{X^2 + 4c_0^2}\right) \quad (2)$$

Here, $c_0$ equals the bulk concentration, and $X$ is a parameter introduced to account for the fixed surface charge, which is given by the following expression:

$$X = \frac{2\sigma}{Fr_p} \quad (3)$$

A full derivation of the model, the values used for the different constants, and the procedure used for fitting the model, are given in the supporting information.

## 3   Results and discussion

### 3.1   Nanopore shrinking

Figure 2 shows a typical sequence of scanning electron micrographs showing the gradual size reduction of a nanopore by local deposition of silicon oxide using a TMOS precursor. The resulting size of nanopores fabricated this way is limited only by the resolution of the SEM. For our experimental setup, nanopores down to 20 nm could be reproducibly fabricated. The shrinking rates measured were ~1 nm/sec for an exposed window of 500 nm x 500 nm, and could be influenced by changing the parameters that influence the electron dose per second that each part of the surface receives. A detailed description of the parameters that influence the shrinking rate can be found in our previous work [15]. The main difference with the previous results is the fact that the shrinking rate does not slow down over time, but remains constant during a series of shrinking experiments because of the nearly unlimited supply of precursor molecules from the sealing device. When using the hydrocarbon contamination in the SEM as precursor, the supply of precursor near the surface is depleted in a relatively short time, while with the sealing device, the used precursors molecules are immediately replaced by new ones, making the process much more stable and reproducible. When the reservoir inside the sealing device is empty, the same slowing in shrinking rate is observed, until after a few minutes, the EBID deposition stops altogether (data not shown).

Local solid-state modification of nanopore surface charges

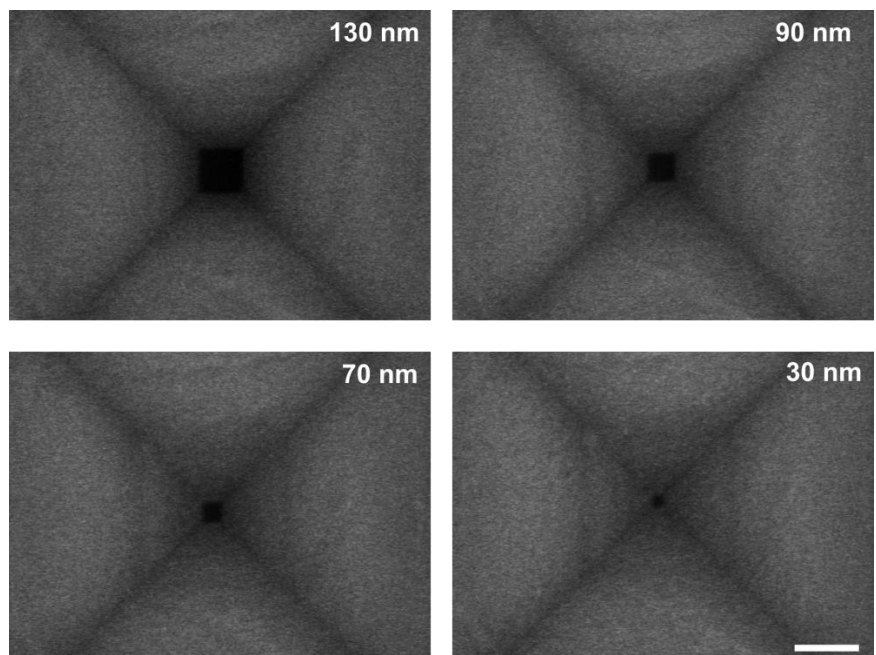

**Figure 2** 4 steps in a nanopore shrinking process, changing the size from 130 nm down to about 30 nm. The scale bar is 200 nm.

Material deposited using EBID typically still contains a reasonable amount of carbon, and therefore differs significantly from the thermal silicon dioxide that originally covers the nanopore surface. The carbon can originate from the incomplete decomposition of TMOS molecules and from the hydrocarbon contamination already present in the chamber of the SEM [20]. Nevertheless, the deposited compound showed a very good resistance to oxygen plasma cleaning, and could be made sufficiently hydrophilic to achieve reasonable nanopore wetting, in contrast to standard nanopores fabricated using the hydrocarbon contamination [15]. It has been reported that the carbon content can be reduced by introducing water together with the precursor, which improves the decomposition [20]. In our case, this may be achieved by introducing one sealing device containing precursor, and another one containing pure water.

*3.2   Concentration dependence of conductance*
To get an idea of the behaviour of a nanopore, and more specifically of the electrical characteristics, we performed a series of ionic measurements for concentrations ranging from 0.1 mM to 1M of KCl. The I-V characteristics and conductance versus concentration behaviour give a good image of the surface properties of the fabricated nanopores. Figure 3a shows the I-V characteristics of a nanopore of ~130 nm. The curves are linear over all concentration regimes, and the pyramid shaped nanopore does not show any ion current rectification, unlike what was observed in funnel shaped track-etched nanopores [24]. This indicates that, as far as ionic conductance is concerned, the nanopore can be considered symmetric, and the model described in the experimental section can provide a good description for the behaviour of the pore. Figure 3b shows the corresponding conductivities as a function of the concentration, and the ionic conductance clearly does not scale linearly with concentration. At higher concentrations, the conductance scales almost linearly with concentration, but below 10 mM, the conductance saturates, and has a value that is much higher than what can be expected from bulk conductance. This effect, commonly referred to as electrical double layer (EDL) overlap, can easily be understood by considering the requirement of charge neutrality inside the nanopore. The bulk of the solution is neutral, but the charges at the surface of the nanopore need to be balanced by an equal amount of counter ions from the solution. Therefore, there

Local solid-state modification of nanopore surface charges will be an excess of counter ions present inside the nanopore. At high concentration, or in other words high bulk conductivity, the counter ions change the conductance only slightly. At low concentrations, however, the counter ions will greatly outnumber the co-ions present inside the pore, and the conductance will depend almost solely on the surface charge of the nanopore. A more detailed description of the phenomenon can be found in literature [21].

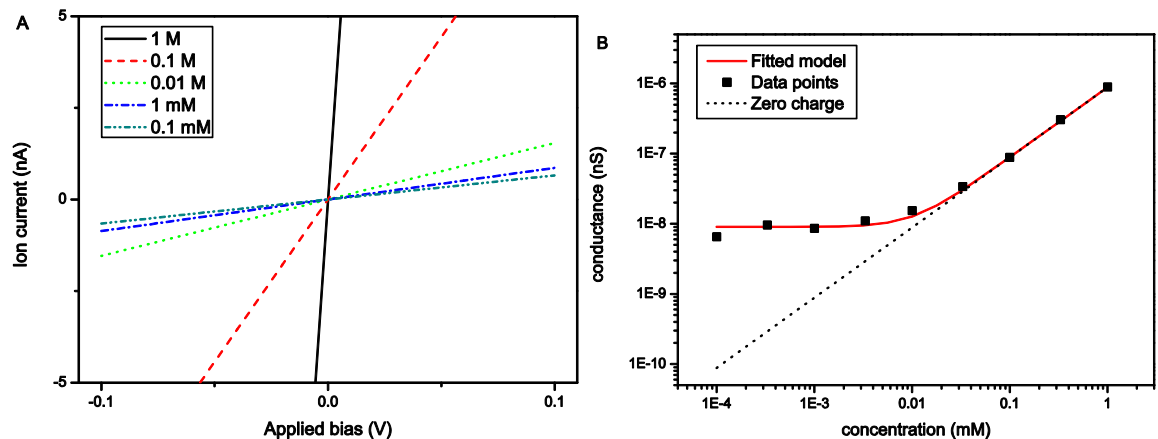

**Figure 3** (a) IV characteristics of a 130 nm nanopore for different concentrations. (b) Corresponding conductance versus concentration plot. The symbols show the measured conductance, the solid line the fitted analytical model, and the dotted line the hypothetical situation without surface charge.

*3.3    Determining the surface charge of a nanopore*

Since the nanopore surface charge is responsible for the observed phenomenon, the shape of the curve actually contains some information on the amount of charge that is present. The solid line in Figure 3b shows a curve which is fitted from the analytical model described in the experimental section. The following fitting parameters were used: pore diameter 130 nm, effective membrane thickness 154 nm, and surface charge 0.16 C/m$^2$. The model fits almost perfectly to the data, indicating that a symmetric nanopore with a constant surface charge is a good approximation for our silicon oxide nanopore. The results were reproducible and showed the same effective membrane thickness for other nanopores fabricated in the same way. This means that for a series of similar nanopores, once the effective membrane thickness is known, the fitting of the model provides an easy way to determine both the pore diameter and the surface charge, which is an important parameter that is normally difficult to determine.

A condition for a good fit between model and data is a surface charge that is constant over the entire range of concentrations. However, an earlier report about the concentration dependence of TEM-drilled nanopores showed a surface charge that decreased at lower concentrations, a phenomenon which was attributed to the chemical reactivity of the nanopore surface, and which could be predicted using a chemical equilibrium model [25]. It appears that the chemical reactivity of the nanopores fabricated using TEM drilling is considerably higher than nanopores fabricated using micromachining and thermal oxidation, possibly due to an increase of chargeable sites caused by the radiation damage of the TEM [8].

*3.4    The effect of shrinking on the surface charge*

Figure 4 summarizes the results of conductance measurement performed at the different stages of the nanopore shrinking experiment shown in Figure 1. Figure 4a shows the I-V curves for each of the shrinking steps, for a concentration of 1 M. It shows a gradual decrease in conductance with each consequent size reduction, like what would be expected intuitively. Figure 4b shows the same sequence of I-V curves, but for a concentration of 1 mM. As can be observed, the conductance first drops significantly after the first shrinking, remains almost constant after the second shrinking, and even increases after the third shrinking step. These counterintuitive observations become much clearer in Figure 4c, which shows the conductance for all concentrations, where each line corresponds to a fitted analytical model. The fitted parameters are summarized in Figure 5, and a few important observations can be made.

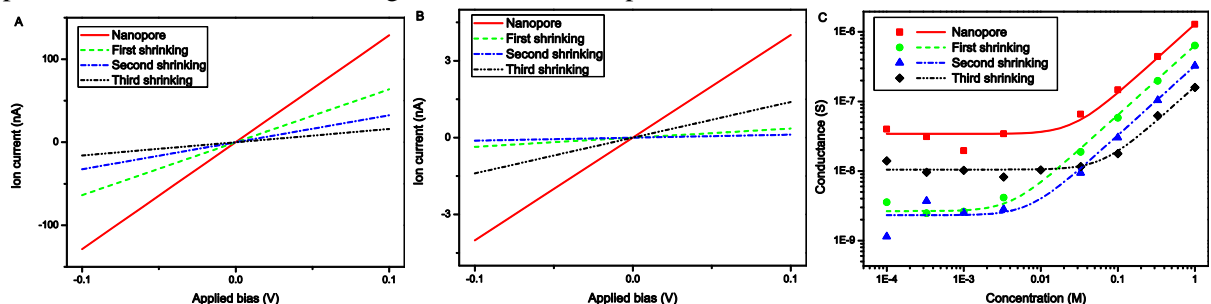

**Figure 4** (a) IV characteristics of a nanopore and 3 sequential shrinking steps for a concentration of 1M. (b) Same IV characteristics for a concentration of 0.1 mM. (c) Corresponding conductance versus concentration, where the symbols show the measurement data, and the lines show the fitted analytical model.

Firstly, the surface charge is constant for each shrinking step over the entire range of concentrations, but it varies significantly with each shrinking step. The charge is reduced after the first shrinking step, slightly increased after the second, and again increased after the final shrinking step. It should be noted, however, that the model is symmetric with regards to surface charge, and gives the same result for a positive and negative surface charge. In other words, the model fitting does not take into account the sign of the surface charge. Silicon dioxide is commonly assumed to have a negative zeta potential and surface charge at neutral pH [14]. Assuming that the surface charge varies monotonously with each shrinking step, this means that the EBID shrinking gradually changes the net surface charge of the nanopore from a negative to a positive charge. This implies a positive surface charge for the EBID silicon oxide at neutral pH. Figure 5 visually represents these assumptions, and presents a summary of the complete shrinking process.

Secondly, the fitted pore diameters do not completely match the values that are measured under SEM. This can of course be explained by the fact that the model assumes a cylindrical shape, which is not true in reality. Moreover, the model uses the same effective membrane thickness, which is valid for nanopores with exactly the same shape. Since the deposition is only performed on a small part of the nanopore, the shape of the nanopore is likely to change with each shrinking step, like visualized in Figure 5.

Finally, these results show that EBID can be used for the local solid-state tuning of surface charge, without the need for surface modification. Figure 6 shows the calculated surface charge as a function of the deposition time, and reveals a near linear relationship. In other words, once this linear relationship is determined for a certain set of deposition parameters, any charge value between the original negative value of -0.163 $C/m^2$ and the final positive value of 0.143 $C/m^2$ can be reached by choosing the appropriate deposition time. This technique can be very useful for nanopore fine-tuning or in any other situation where local surface charge control is desired.

Local solid-state modification of nanopore surface charges

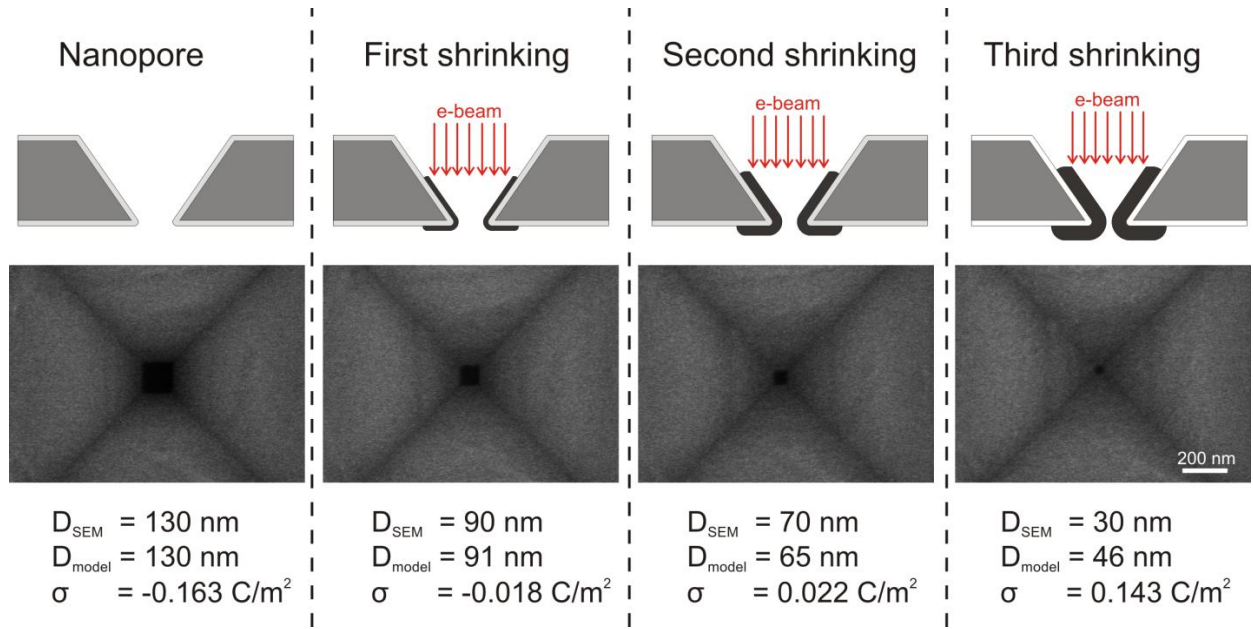

**Figure 5** Summary of the shrinking process, with a schematic drawing of the expected EBID process, the nanopore size estimated from both SEM and model, and the surface charge together with the postulated sign.

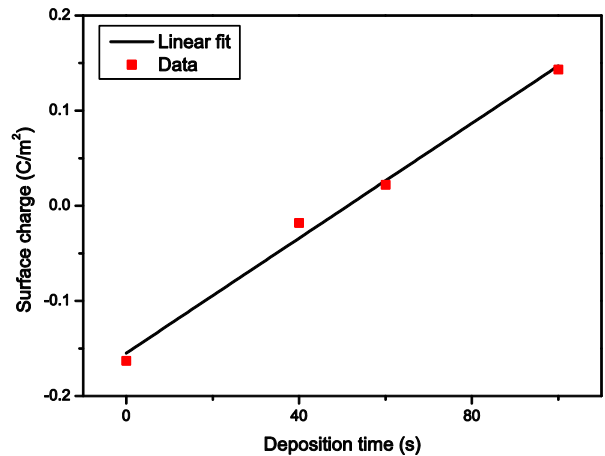

**Figure 6** Calculated surface charge as a function of deposition time.

### 4  Conclusions

This work has shown the electrical characterization of artificial nanopores that have been resized using electron-beam induced deposition (EBID) in an unmodified SEM, using a sealing device for the release of TMOS precursor. Measurement of the ionic conductance as a function of salt concentration, together with an analytical model, has provided insight on the amount and the stability of the surface charges present inside the nanopore under different salt concentrations. It was shown that for both the unmodified nanopores, which have a thermal silicon oxide surface, and the nanopores shrunk by EBID, the surface charge is constant over the entire range of concentrations. It was also postulated that the charge of the material deposited by EBID has a different sign than the original nanopore surface, causing the net surface charge to vary with each consequent shrinking step.

These results suggest that EBID in an unmodified SEM provides a cheap and readily available technique to produce nanopores of sizes of only a few nanometres, where the surface properties can be modified by

Local solid-state modification of nanopore surface charges choosing different precursors and deposition parameters. When the surface charge is constant over a wide range of concentrations, which is the case for both thermal and deposited silicon oxide, combining concentration dependent measurements with the fitting of an analytical model can provide an easy and reproducible way to estimate the surface charge, thus allowing researchers to quickly choose the nanopore that is most suited for the envisioned application.